\documentclass[twocolumn,twoside,slac]{revtex4}
\usepackage{graphicx}
\usepackage{fancyhdr}
\begin{document}

\title{The FRED Event Display: an Extensible HepRep Client for GLAST}
\author{M.Frailis}
\author{R.Giannitrapani} 
\affiliation{Dipartimento di Fisica, Universit\`a degli Studi di Udine - Italy}

\begin{abstract}
  A new graphics client prototype for the HepRep protocol is presented. Based on
  modern toolkits and high level languages (C++ and Ruby), Fred is an experiment
  to test applicability of scripting facilities to the high energy physics event
  display domain. Its flexible structure, extensibility and the use of the
  HepRep protocol are key features for its future use in the astroparticle
  experiment GLAST.
\end{abstract}

\maketitle 
\thispagestyle{fancy}
\section{Introduction}

Event display is becoming more and more a key component of modern high
energy physics and astroparticle experiments (see \cite{Drevermann} for a nice
introduction on the idea of event display); thanks mainly to the wide
availability of powerful desktop computers and to the performances of modern
programming languages, we have now the possibility to exploit a full 3D event
display with modern GUIs and users interactivity at a level that was almost
impossible few years ago.

FRED (that is an acronym for {\em FRED is a Recursive Event Display}) is part of
the ongoing efforts for developing an extensible event display
framework\footnote{For a description of this general framework see the companion
  talk \cite{heprepglast}.}  for GLAST \cite{glast}, an international
astroparticle experiment lead by NASA; GLAST is an orbital gamma ray telescope
that will measure photons in a broad range (from $20$ MeV to more than $300$
GeV) not covered at the moment by other similar instruments. Its structure,
mainly derived from HEP experiments, consists of a standard silicon strip
tracker, an hodoscopic calorimeter (CsI) and a plastic scintillator
anticoincidence. The launch is scheduled for late 2006 and the lifetime will be
at minimum 5 years. Both events topology and detectors geometry in GLAST are
rather simple with respect to typical modern HEP experiments; nowadays its user
requirements for event display are almost the same, i.e. a fast, modular,
extensible, flexible application with a suitable GUI. Modularity is particularly
important for GLAST since the development of a modern event display has started
late, and almost all the infrastructure of the offline software have been
already decided and mostly implemented; in particular GLAST is adopting the
GAUDI\cite{gaudi} framework for the offline software.  With this respect it is
important that the new event display does not force changes to the already
finalized choices of the software group; on the other side it is quite important
that it does not rely heavily on the offline software infrastructure that it is
prone to changes that should not affect the graphics representation.

At the moment GLAST has already a GUI service that, althought fast, easy and
fully integrated in the software framework, misses some of the structural
requirements, expecially on the interactivity side; for example it is not
possible to inspect graphics representations for physics attributes.

\section{FRED and HepRep}

To start to use a new external program can be risky in an andvanced state of the
infrastructure software (framework, montecarlo, datastores, geometry repository
etc etc); this leaded quite naturally to a server/client
paradigm\footnote{Please note that here server/client have architectural
  meanings; in a real implementation they can be two remote applications as well
  as childs processes of a single local application.} in order to separate the
physics issues from the graphics ones and to have minimum impact on the offline
software of GLAST. The main idea, borrowed by other event displays and
expecially from WIRED\cite{wired}, is to encapsulate the graphics specific
issues in a client program, leaving to a server (that for GLAST lives in GAUDI)
all the physics related issues that are specific to the experiment, like for
example the event structure, relevant physics attributes that need to be exposed
to the event display user, geometry of the detectors etc. 

Instead of working out a new possible protocol for our framework, we decided to
adopt HepRep\cite{heprep}, a well designed protocol whose generic design allows
for a lot of flexibility without enforcing too tight choices to the experiment
software; althought it has been originally designed and implemented for WIRED,
it is completely general and application independent.  Its abstract
and pluggable tree structure is able to accomodate all the specific issues
of an experiment and is really easy to be customized and to be extended.

FRED is just a new HepRep client; started as a specific GLAST application,
thanks mainly to the use of HepRep and some architectural choices described
later, it is now mainly experiment independent.  In common with WIRED, two kinds
of HepRep sources have been implemented in FRED, i.e. a CORBA\cite{corba} source
for client/server remote or local operations and an XML persistency file
mechanism. Thanks to this, WIRED and FRED will be completely interchangeable for
the GLAST experiment, reducing in this way the degree of commitment of the
experiment to a single product.  Moreover it is possible to plug in other
possible sources both as network protocols or persistency file formats (for
example WIRED already support Java RMI) to cope with experiment specific
requirements.

\begin{figure}[t]
\includegraphics[width=80mm]{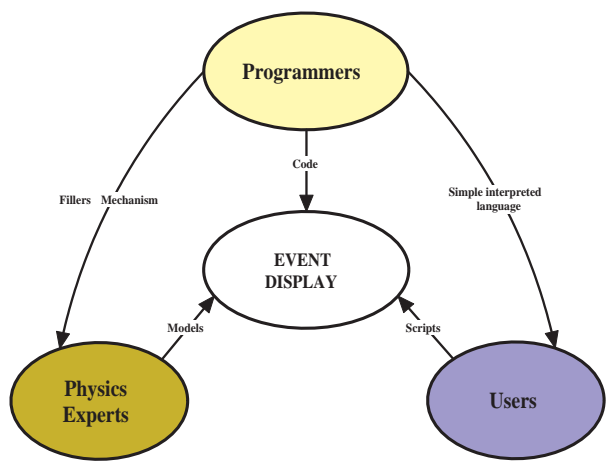}
\caption{The FRED paradigm.} \label{event}
\end{figure}

The innovation of FRED rely in the implementation of an open and extensible
architecture that enables users to customize, expand and evolve the main
program. In its development we tried to separate the roles of the three main
actors in an event display application (see figure \ref{event}): 

\begin{description}
\item the {\bf\em programmers} (normally graphics experts) that design and
  implement the main programs and all the infrastructure code.
\item The {\bf\em physics experts} who decide what to represent, i.e. the {\em
    models}, and what physical information will augment the graphical
  representations of an event; for example in GLAST we are providing a
  ``fillers'' mechanism, fully described in \cite{heprepglast}, that can be
  easily provided by physicists to fill the HepRep hierarchy starting from the
  transient or permanent data stores for both the event content and the geometry
  structure.
\item The {\bf\em end users}, who need a fast and easy way to customize the
  application to their needs, for example augmenting the GUI by adding new
  buttons, menus voices, general widgets, color schemes etc., or providing
  completely new features like new input/output formats or experiment specific
  functionalities. Typical ways to do this is via compiled plugins and/or
  scripting capabilities.
\end{description}

In this respect FRED can be thought of more as an event display framework
rather than as an application.

\section{Implementation issues}

After a first prototype of FRED implemented completely in C++ in late 2002, we
decided, following the discussion in the previous section, to step over to a
more open architecture that is able to expose to the end user most of the
internal interfaces of the event display. The main reason is to give users the
possibility to customize mostly all FRED aspects, from the GUI to the internal
functionalities and the interactivity of the user. To this extent the use of a
scripting language is quite natural, but for good performances on an intensive
graphics application like an HEP event display, some parts have to be desgined
and implemented in C++; FRED has been designed as a mixture of these paradigms
and the bridge between the two sides (compiled and interpreted one) have been
implemented thanks to SWIG\cite{swig}, a semi-automatic generator of
scripting languages extensions.

\begin{figure*}[t]
\includegraphics[width=120mm]{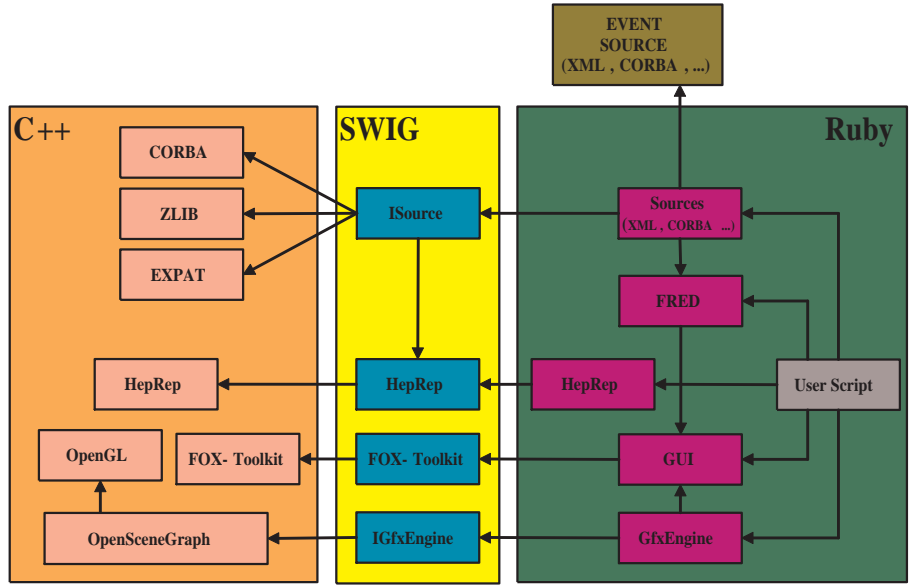}
\caption{The FRED structure.} \label{newFRED}
\end{figure*}

A simplified scheme of the main FRED components, shortly described in the next
sections, is depicted in figure \ref{newFRED}.

\subsection{Scripting and GUI domain}
For the scripting language, after some comparative evaluations, we have decided
to use RUBY\cite{ruby}, a highly dynamic object oriented scripting language; the
choice derived mainly from the simplicity and elegance of the language, together
with a rather mature development of its main features and of usable libraries
that cover all our needs. Moreover it is really simple to provide extensions to
the language from shared C++ libraries (using also SWIG) with a negligible
impact on performances and a very small footprint in both memory and disk space.

For the GUI we have adopted the FOX-toolkit \cite{fox}, a C++ multiplatform,
open source, extensible, clean library that is gaining popularity over other
similar toolkits (Tk, GTK, QT etc.). This toolkit provides out of the box many
standard widgets like toolbars, menus, tooltips, trees, shutters, canvases
(comprising an OpenGL one) etc. Moreover, thanks to its clear C++ design, its
possible to easily subclass existing widgets to customize them or to create new
ones from scratch.

Since a RUBY wrapping of that library has already been developed \cite{fxruby},
it has been very easy to build our framework in a way to let users add new GUI
components and new widgets\footnote{For this we wish to thank the FOX and Ruby
  communities (expecially Lyle Johnson) for a lot of helpful advices and for
  providing us a ``user support'' that can teach something to propetary
  software.}.  This is possible both with scripts plug-in loaded at startup or
interactively during a live session via an interactive Ruby interpreter.

\subsection{C++ domain}
As already noted, most of the computationally heavy parts of the event display
have been implemented in C++. 

\begin{itemize}
\item All the IO features (expecially related to reading and writing of
  compressed XML files) have been implemented in C++ using various public
  available multiplatform libraries \cite{expat, zlib}. 
\item For the graphical representation of the event we choose OpenGL
  \cite{opengl}, that is de facto the industry standard for both 2D and 3D
  graphics (and that work transparently with and without dedicated hardware); in
  particular we are using at the moment OpenSceneGraph \cite{openscene}, that is
  an open source scene graph manager very helpful to augment performances and to
  use lot of already disposable and efficiently implemented computer graphics
  algorithms (like text representations, views culling, depth sorting of
  representables tree etc etc). For the future we are planning to develop our
  own HepRep based scene graph to minimize duplication of in memory information.
  For performance reasons all the OpenGL calls have been encapsulated in the C++
  side, with no Ruby wrapping at the moment.
  
\item We decided to keep the full HepRep hierarchy, that potentially can be very
  big for a typical event in GLAST, in the C++ domain; this have been proved to
  help performances in both creation in memory and access for information
  retrieval. The full hierarchy is anyway wrapped at Ruby level in order to let
  user'scripts acess it.
\item The CORBA connection has been implemented in C++ by using
  ACE/TAO\cite{ace}, a fast, multiplatform and stable open source implementation
  of the CORBA standard. It is provided as an optional shareable library
  component that can be loaded at startup by Ruby.
\end{itemize}

\section{Some preliminary results}

FRED has been released as a GLAST internal beta in January 2003; the main
planned features are present in this preliminar delivery and provided good
performances and interaction results:

\begin{figure*}
\centering
\includegraphics[width=135mm]{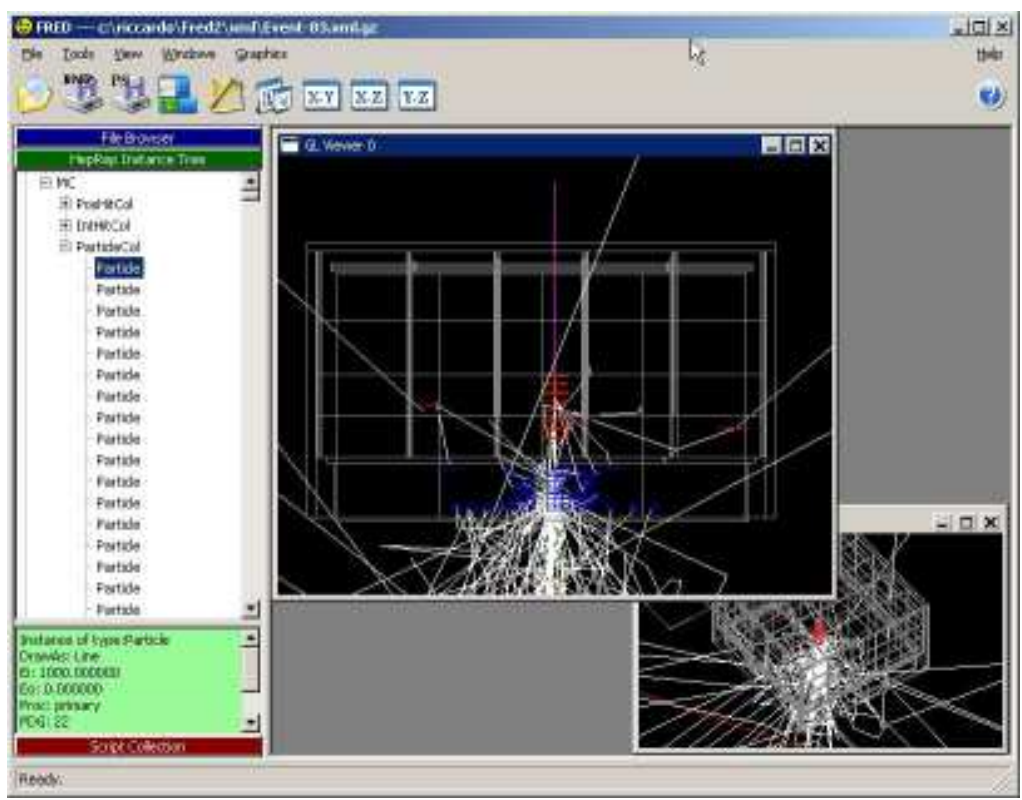}
\caption{A GLAST Montecarlo event and geometry shown in the FRED gui.} 
\label{fred}
\end{figure*}

\begin{description}
\item {\bf Multiplatform}: tested and supported on Windows 2000/XP and Linux RH,
  but should work on many Unix flavours thanks to the multiplatform nature of
  all the toolkits adopted by FRED.
\item {\bf HepRep Sources}: in this version the user is able to load compressed
  HepRep XML files (althought not all the possible HepRep shapes have been
  already implemented, just the GLAST related ones) or connect to a test CORBA
  server implemented in the GLAST framework (i.e. GAUDI).  
\item {\bf Output}: both raster (BMP, PNG and JPG) and vectorial (PostScript)
  formats are supported.
\item {\bf Interaction}: usual zoom, pan, rotate operations via mouse and keys;
  it is possible to browse the HepRep instances tree and access all the
  information attached to graphical representables in the event, both
  interacting with a tree widget or directly selecting the object from the 3D
  view.
\item {\bf Graphics}: using OpenGL, FRED automatically uses any available
  hardware acceleration; for simple events (like GLAST ones) and using a
  wireframe representation, performances are very good also in software mode on
  a typical desktop machine configuration.
\item {\bf Scripting}: script API gives access to all the main functionalities
  of FRED, comprising the possibility to add new widgets to the GUI. FRED has a
  simple internal editor for scripts and also an interactive embedded
  interpreter.
\end{description}

In figure \ref{fred} we depicted a typical GLAST montecarlo event in the FRED
GUI (where it is possible to see the independent multiple views capability of
FRED) More screenshots and information can be found on the web page of FRED
\cite{fredweb}.

\section{Outlook and conclusions}

We are actually working to finalize a stable release and to add some new
features;

\begin{itemize}
\item New HepRep sources, expecially an {\bf HTTP} one for browsing events
accessible from internet

\item Use of the Ruby RMI mechanism, {\bf druby}, for remote control of FRED
from other applications 

\item Export various format for photorealistic rendering of the event (we are
working to a {\bf POV} exporter and to a {\bf RenderMan} one)

\item More than just wireframe graphics (filled volumes, semitransparency,
  outlines etc etc); in particular we are working to a layer mechanism in OpenGL
  that can provide more traditional layered 2D views together to 3D z-buffered
  wireframe ones.

\item A ``batch'' mode to produce event images (vectorial or raster) without
starting the GUI

\item Options panel, with possibility to save preferred configuration from one
session to the other

\item More GLAST specific features (that will be collected in a single plugin),
  like for example the possible back interaction with GAUDI and the physics
  algorithms. 
\end{itemize}

Althought FRED is an ongoing project and still lot of work have to be done to
provide a stable, full fledged event display suited for modern high energy and
astroparticles experiments, the prototype experiments and the overall
architecture seem encouraging. In particular we think that the easiness with
which it is possible to customize and extend FRED, also interactively during a
working session, will provide GLAST and other potential users an interesting way
to fine tune the event display to their specific needs.

\begin{acknowledgments}
  We wish to warmly thanks Joseph Perl and Mark D\"onszelmann for helpful discussions
  on event display and related issues and for providing us HepRep and WIRED;
  FRED had never started without them. We wish to thanks also all the GLAST
  software group for encouraging us on our way. 
\end{acknowledgments}

\end{document}